# High-entropy van der Waals materials formed from mixed metal dichalcogenides, halides and phosphorus trisulfides

Tianping Ying[1*], Tongxu Yu[2], Yu-Shien Shiah[1], Changhua Li[3], Jiang Li[1], Yanpeng Qi[3*], Hideo Hosono[1*]

1. Materials Research Center for Element Strategy, Tokyo Institute of Technology, Yokohama 226-8503, Japan
2. Laboratory of Advanced Materials, Fudan University, Shanghai 200438, China
3. School of Physical Science and Technology, ShanghaiTech University, Shanghai 201210, China

**ABSTRACT:** The charge, spin, and composition degrees of freedom in high-entropy alloy endow it with tunable valence and spin states, infinite combinations and excellent mechanical performance. Meanwhile, the stacking, interlayer, and angle degrees of freedom in van der Waals material bring it with exceptional features and technological applications. Integration of these two distinct material categories while keeping their merits would be tempting. Based on this heuristic thinking, we design and explore a new range of materials (i.e., dichalcogenides, halides and phosphorus trisulfides) with multiple metallic constitutions and intrinsic layered structure, which are coined as high-entropy van der Waals materials. Millimeter-scale single crystals with homogeneous element distribution can be efficiently acquired and easily exfoliated or intercalated in this materials category. Multifarious physical properties like superconductivity, magnetic ordering, metal-insulator transition and corrosion resistance have been exploited. Further research based on the concept of high-entropy van der Waals materials will enrich the high-throughput design of new systems with intriguing properties and practical applications.

## 1. INTRODUCTION

The isolation of graphene by a 'scotch-tape' method ushers in an era of intensive research in van der Waals (vdW) materials[1]. Different from the other three-dimensional (3D) counterparts, vdW materials have weak interlayer interaction, and thus hold unique tuning knobs such as stacking, twisting and intercalation flexibility[2]. Stacking from monolayer to multilayers turns out to be an effective way to modulate the band structures and the strength of charge/spin correlations[3,4]. The thus-obtained atomic-thin heterostructures hold the promise to be next-generation ultra-compact electronic/spintronic devices[5]. Further introducing chemical species into the interlayer space of vdW materials will alter the crystal structures and valence states. Reversible intercalation and deintercalation cycles are crucial to realize broad applications in energy storage and conversion[6,7]. The modification flexibility and intriguing characteristics of vdW materials have made them the rising materials in the broad research community. Very recently, the tunable transformation among semimetal, Mott insulator and superconductor through the "magic angle" twisting of bilayer graphene brings a new storm to condensed matter physicists[8,9]. In the scenario concerning 2D exploitation, the discovery of new vdW materials is always the main theme, e.g. the MXene family further expands the 2D material candidates to several dozens[10].

The concept of high-entropy alloy (HEA), proposed at nearly the same time as the discovery of graphene, is a kind of alloying strategy that involves the combination of multiple principal elements in relatively high concentrations[11,12]. Simply put, HEA requires the configurational entropy ($T\Delta S_{mixing}$) contributing to the free energy overwhelming the formation enthalpy ($\Delta H_{intermetallic}$). This requirement could be further put into a rough criterion of alloying five or more elements each with concentration range of 5~35 at.%. Under this constraint, there are basically infinite possible combinations with highly tunable charge, spin and composition degrees of freedom[13]. Furthermore, HEA and its extension to various 3D ceramics are well known for their excellent mechanical performance and outstanding resistance to degradation, which could be valuable for employment under harsh environments[14–19]

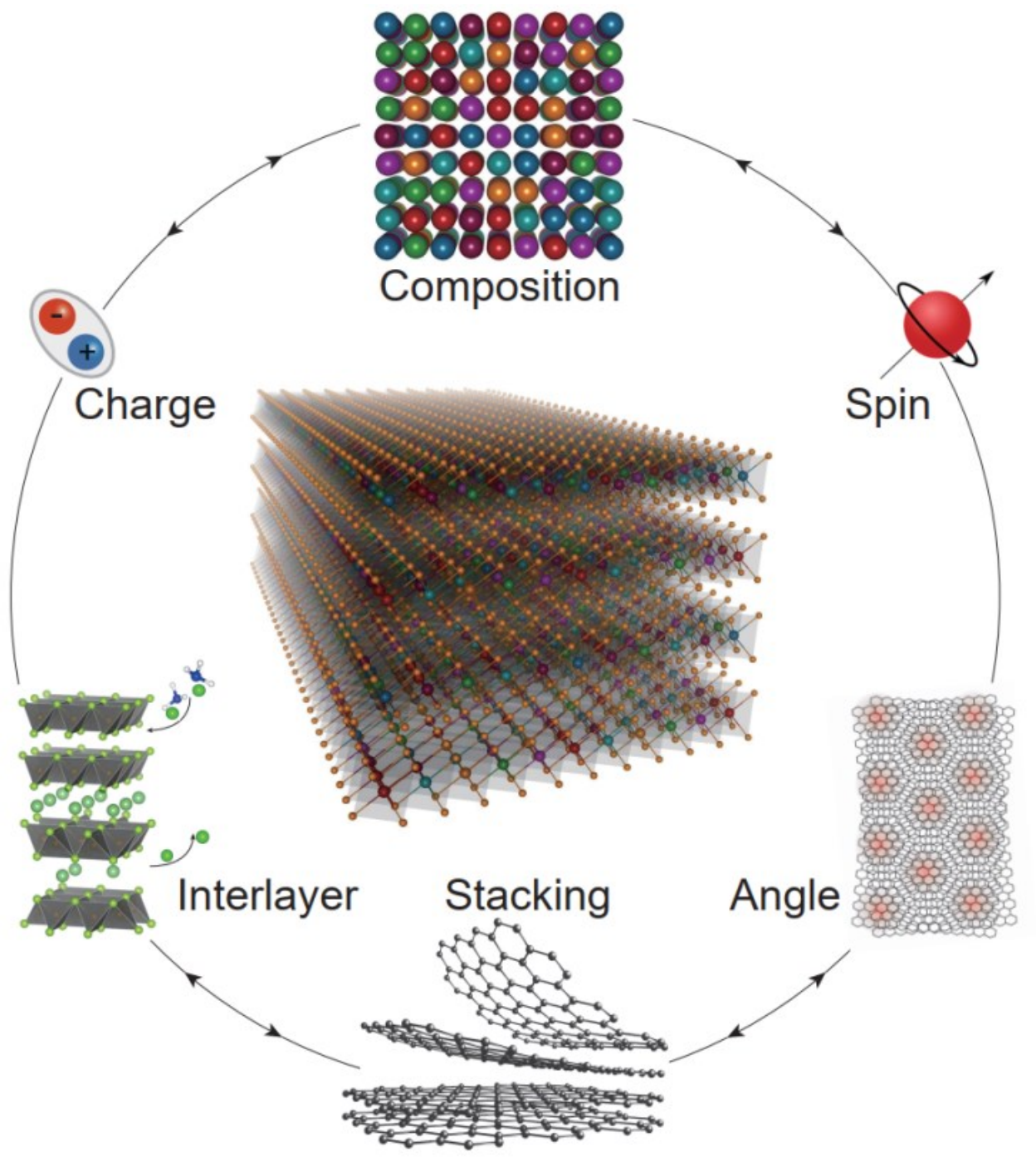

**Figure 1.** Design concept of HEX materials. The charge, composition and spin degrees of freedom inherited from HEA and interlayer, stacking and angle degrees of freedom ascribed to vdW materials can be intermingled together into a brand-new material category, namely HEX materials.

With the impetus to blend the characteristics of these two categories of materials into a single system, here we elaborate on the design, synthesis, and characterization of a new family of vdW materials, namely high entropy van der Waals materials (HEX). As a first attempt to realize HEX materials (**Figure 1**), we start from the already known vdW materials with similar crystal structures, for example transition metal dichalcogenides, halides or phosphorus trisulfides. A batch of single-phase HEX materials have been successfully acquired, which can be exfoliated and intercalated with ease. Given the abundant elemental combinations and the layered structure, bountiful physical properties have been inherited or emerging, some of which are elaborated here.

## Experimental

Polycrystalline HEXs were synthesized using the conventional solid-state method. Powder elements with their nominal composition were weighted, thoroughly grounded, and pressed into pellets. $HECl_2$ series are prepared by using constituent $M$$Cl_2$ raw materials. The samples were sealed into quartz tubes and slowly raised to the desired temperature (see Table 1 for details) and hold for 24-48 hours, which were followed by quenching into ice water or furnace cooling. Single crystalline HEXs were grown by using chemical vapor transport (CVT) method with suitable temperature gradients for different compositions (Table 1). After keeping at high temperature for 2-30 days, the single crystals could be harvested by quenching into ice water. Iodine was used as the transport agent while Se can also be used for the growth of $HESe_2$. We note that when $T\Delta S_{mixing}$ and $\Delta H_{intermetallic}$ are close, quenching from high temperature is necessary to avoid the spinodal decomposition. Based on our discovery of 17 HEX materials in the present work, it is the comparable electronegativity rather than the ionic radii that permit the acquisition of the single phase.

X-ray diffraction (XRD) patterns were acquired using a Bruker D8 Advance diffractometer with Cu-$K_\alpha$ radiation (wavelength $K_\alpha 1$ = 1.54056 Å, $K_\alpha 2$ =1.544390 Å) at room temperature. The FullProf software package was used for the Rietveld refinement. The composition mapping of single-crystalline (Ti,V,Cr,Nb,Ta,)$Se_2$, (Mn,Fe,Co,Ni)$PS_3$ and (V,Cr,Mn,Fe,Co)$I_2$ were determined using an electron-probe microanalyzer (EPMA, JEOL, Inc., model JXA-8530F). The compositions of the rest samples were carried out by using an energy-dispersive X-ray spectrometer coupled with a scanning electron microscope. Topographic images were taken by using an atomic force microscope (AFM, Bruker, multimode8-FSCamp12). Transport, magnetic, and thermodynamic properties were measured using a physical property measurement system (PPMS, Quantum Design) and a superconducting quantum interference device (SQUID) vibrating sample magnetometer (SVSM, Quantum Design). The chemical state of the elements was investigated by using X-ray photoelectron spectroscopy (XPS, Thermo Fisher 250Xi). Corrosion resistance measurements were carried out in acid (8.1M $HNO_3$, room temperature, 10 minutes), base (0.5M NaOH, room temperature, 10 minutes) and organic solution (1.0 mmol of butylamine mixed in 2 ml tetrahydrofuran, 40 °C, 1 hour). The reactions with organic solution were conducted in stainless steel autoclave fitted with a glass mantel and a magnetic stirrer. The autoclave was evacuated at first, and then charged with 0.5 MPa $O_2$ and 0.5 MPa CO. The concentrations of the dissolved Se were determined by inductively coupled plasma (ICP). Nitrogen adsorption/desorption measurements were performed to estimate the Brunauer-Emmett-Teller surface area of the as-synthesized polycrystals (BELSORP-mini II, BEL).

Few-layer HEXs can be easily exfoliated by scotch tape and transferred to $SiO_2$/Si wafer by using the Polydimethylsiloxane (PDMS) method under optical microscope. Monolayer $HES_2$ had been acquired through $Al_2O_3$-assisted exfoliation method[20]. Intercalation of HEXs was carried out by using the standard liquid ammonia method described in Ref. 21.

### Results and Discussion

The crystal structures of HEX families share a common layered nature (**Figure 2**a-d). The HE sites are occupied by randomly distributed principal metals, while for each layer, it is terminated by X with weak vdW interaction in between. Single crystals of several millimeters can be easily acquired by chemical vapor transport method (left panels of **Figure 2**e-h). The distribution homogeneity of the elements is confirmed by the mapping of electron probe microanalyzer and Energy Dispersive X-Ray Analysis (**Figure 2**i-l and Supporting information). All the peaks of the x-ray diffraction pattern (**Figure 2**m-p, red curves) can be well indexed (blue vertical bars), and the single crystalline diffraction patterns (black curves) show clear preferred orientation perpendicular to the *ab*-plane. Their Rietveld refinement results can be found in Supporting information. **Table 1** included an extra dozen HEX materials discovered and characterized in this paper, from which we highlight that it is not necessary to require all the individual components to share the same crystal structure. For example, $(Ti,V,Cr,Nb)_{0.8}(Fe,Mn)_{0.2}Se_2$ could tolerate up to 20 at.% isomerous alloying of (Fe/Mn), while Fe-Se and Mn-Se alone do not possess the layered $CdI_2$-type structure. This observation could further increase the flexibility of composition choice for material design.

As a prominent feature of vdW materials, HEX single crystals could be easily exfoliated into few layers by conventional scotch tapes (right panels of **Figure 2**e-h). The iridescent color is optical interference caused by different layers of the thin films on the $SiO_2$/Si substrate. Their topographical landscapes (AFM images) are shown in **Figure 2**q-t. From the cross-sectional profile of (Mn,Fe,Co,Ni)$PS_3$ we can see that the thinnest area achieved so far is 2.5 nm, roughly the height of four stacking layers. Monolayer $HES_2$ has been acquired by using $Al_2O_3$-assisted exfoliation method[20] (Figure S45,46).

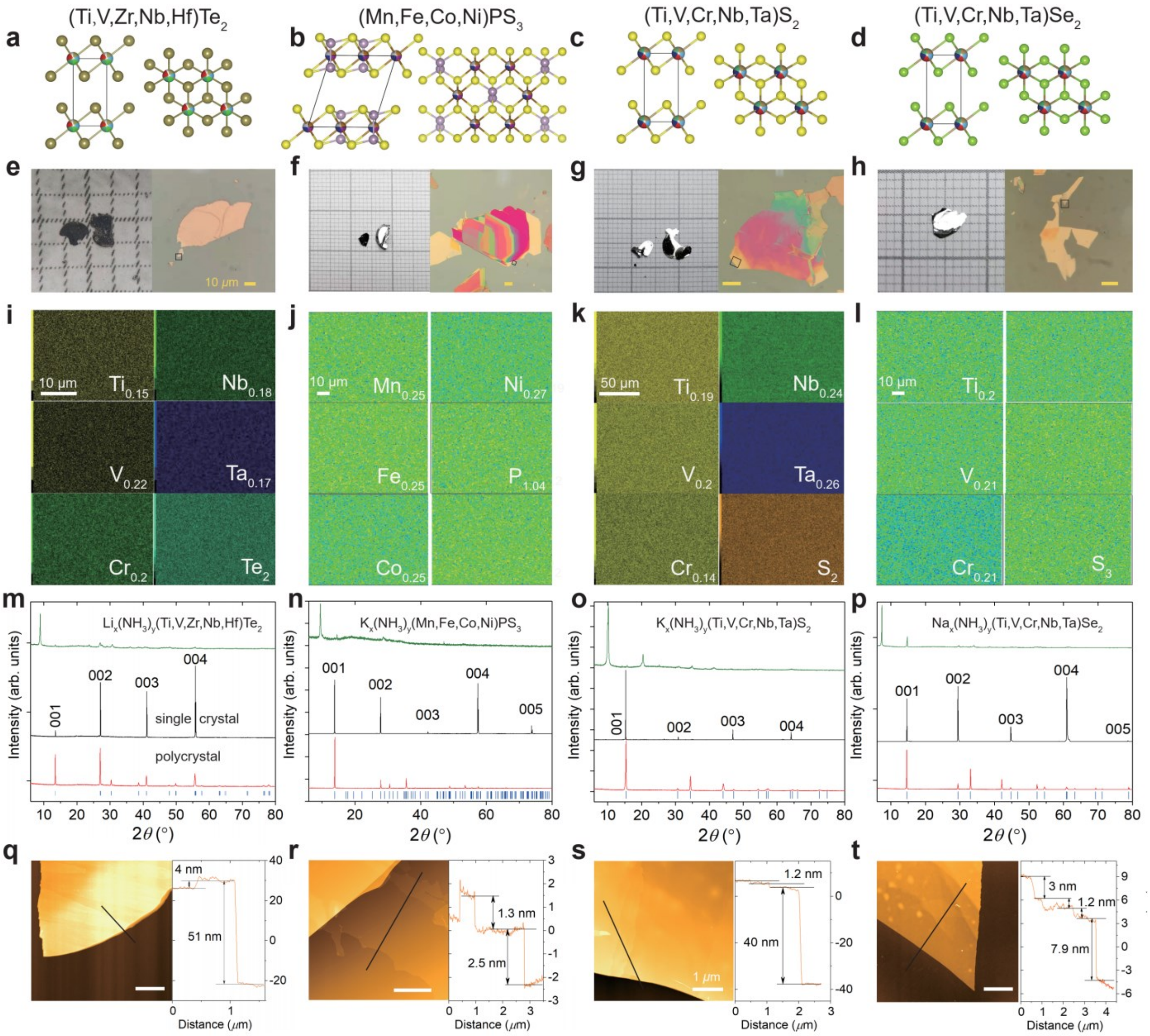

**Figure 2.** Structure, exfoliation and intercalation of HEX materials. **a-d** Crystal structure of HEX. Left and right panels are the projection along *a*- and *c*-axis, respectively. The high-entropy metallic sites are occupied by transition metal atoms (represented by multi-color balls) while the interlayers are terminated by chalcogen or halogen species such as S, Se, Te, or $PS_3$ with week vdW interaction in between. **e-h** Optical images of HEX single crystals on a millimeter paper and the exfoliated few layers on $SiO_2$/Si wafer. **i-l** Quantitative element mapping of HEX by EDX (**i, k**) and EPMA (**j, l**). **m-p** X-ray diffraction patterns of polycrystalline (red), single-crystalline (black) and metal-intercalated (green) HEX materials. **q-t** Atomic force microscope (AFM) topographic image of the exfoliated HEX thin films (left panels of **e-h**) and their cross-sectional profile along the black line.

Another unique characteristic of vdW materials is the ability to be intercalated by the vapor of alkali metals[22], solid-state gating[23,24] or assisted by solvents such as liquid ammonia[21,25]. The increased lattice constants evidenced by the shift of the peaks towards lower angles (**Figure 2**m-p) imply the successful intercalation of the alkali metals, possibly together with ammonia/amide species according to the large expansion of the *c*-axes (50% for Li doped $HETe_2$, 40% for K doped $HEPS_3$, 51% for K doped $HES_2$ and 22% for Na doped $HESe_2$). We also found a time-dependent structure evolution in some cases once the samples were removed from the liquid ammonia environment, which is reminiscent of the rich structures in metal-intercalated FeSe with the variation of surrounding ammonia[26]. A systematic evolution of the valence states from raw $HES_2$ to monovalent (K) and divalent (Ba) intercalation can be observed from the continuous peak shift of XPS (**Figure S47**a,b) for S and Ta, while the difference between mono- and divalent intercalation is not apparent for the remaining transition metal elements (**Figure S47**c-f).

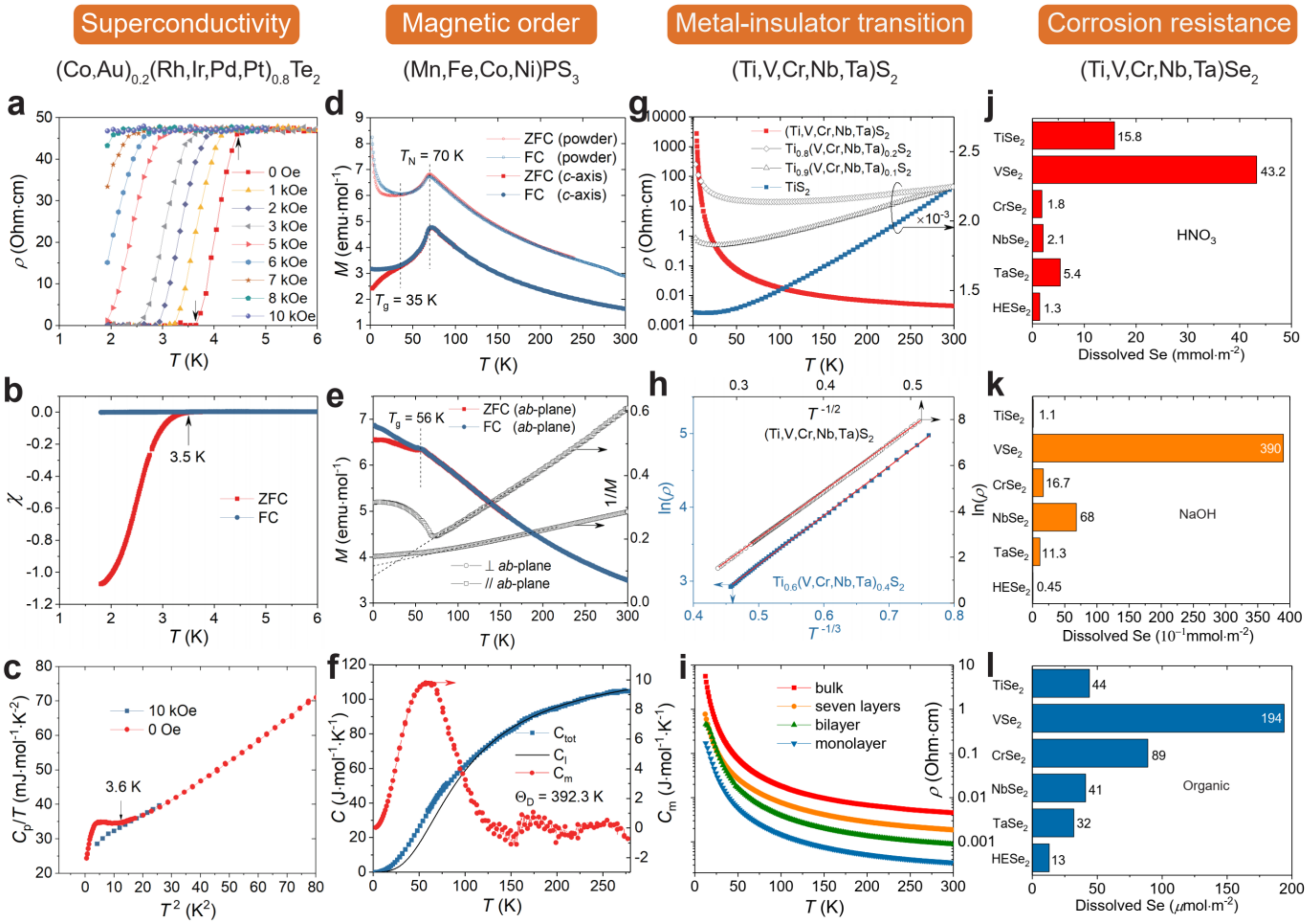

**Figure 3. Physical properties and potential applications of HEX materials. a-c** Resistivity, magnetization and heat capacity of $(Co,Au)_{0.2}(Rh,Ir,Pt,Pd)_{0.8}Te_2$. **d,e** Magnetization of $(Mn,Fe,Co,Ni)PS_3$ with the external magnetic field ($H$ = 500 Oe) perpendicular to or along the *ab*-plane. The antiferromagnetic transition ($T_N$) and spin glass transition temperature ($T_g$) are indicated. The results of the polycrystalline sample are also included. **f** Heat capacity of $(Mn,Fe,Co,Ni)PS_3$ from 2 to 280 K. The magnetic contribution (red) is extracted from subtracting the lattice specific heat (black) by fitting the Debye formula above 160 K. **g** Metal-insulator transition of $(Ti,V,Cr,Nb,Ta)S_2$ with the variation of the composition. **h** Two-dimensional variable-range hopping of the resistivity in $(Ti,V,Cr,Nb,Ta)S_2$ system. The change of the linear dependence of ln($\rho$) vs. T from $T^{-1/3}$ to $T^{-1/2}$ indicates the opening of a Coulomb gap at low temperature. **i** Thickness dependent resistivity of $(Ti,V,Cr,Nb,Ta)S_2$ from 12 to 300 K. **j-l** Corrosion resistance measurements of $(Ti,V,Cr,Nb,Ta)Se_2$ and its individual components against acid ($HNO_3$), base (NaOH) and organic (butylamine mixed in tetrahydrofuran) reagents.

The multiple degrees of freedom and the infinite combination in HEX materials provide us a unique opportunity to explore multifarious physical properties. Here we give an example of $(Co,Au)_{0.2}(Rh,Ir,Pd,Pt)_{0.8}Te_2$ which shows superconducting transition temperature ($T_c^{onset}$) at 4.5 K and a bulk response at 3.6 K (**Figure 3**a-c), comparable to the highest $T_c^{onset}$ in the metal-ditelluride systems.[27,28] Nevertheless, it should be emphasized that the maximum $T_c$s in these ternary systems usually appear at the boundary of structural instability or on the verge of $Te_2$ dimer breaking. And their superconducting regions are relatively narrow,[27–31] distinct from the robust and easily accessible superconductivity in $HETe_2$. The natural absence of structural transition and the widespread dimer fluctuation play an important role, which can be viewed as a cocktail effect of the high entropy. Further enhancement of the $T_c$ is possible by modulating the composition which directly influences the valence state as well as the strength of electron-phonon coupling. An apparent suppression of the superconductivity by increasing the doping content of Co can be observed from $Co_{0.1}(Rh,Ir,Pd,Pt)_{0.9}Te_2$ to $(Co,Rh,Ir,Pd,Pt)Te_2$ (**Table 1** and Supporting information). Deliberate incorporation of elements with local spin state can also lend us a platform to investigate the unconventional pairing mechanism of superconductivity and its ostensible competition with magnetism.

With elements bearing different local moments included, HEX materials are anticipated to have rich magnetic behavior. Take $(Mn,Fe,Co,Ni)PS_3$ for example (**Figure 3**d-f), one can see a clear AFM transition at $T_N$ = 70 K. With lowering the temperature, a spin glass state emerges at $T_g$ = 35 K, as evidenced by the bifurcation of the field cooling (FC) and zero-field cooling (ZFC) curves. However, the frozen temperature along the *ab*-plane shifted to 56 K and the AFM transition is almost indistinguishable. A broad hump in the heat capacity of $HEPS_3$ at ~60 K is evidenced after deduction of the lattice contribution which hints a typical Heisenberg-AFM behavior[32]. The observed AFM and different reentrance temperatures of the spin-glass states along different directions are much richer than the mono-metallic phosphorus trisulfides, where each compound shows a certain angle-independent transition temperature[32–34]. **Table 1** listed several other HEX materials with AFM or spin-glass ordering. Multifarious magnetic configurations are waiting to be explored, among which ferromagnetic (FM) HEX materials are of paramount importance[20,35,36]. Considering the rareness of ferromagnetism in 2D materials, the richness of HEX materials provides promising opportunities. Detailed magnetic characterization, e.g., neutron scattering, on these systems is indispensable to understand the multi-magnetic transition temperatures and the complicated mutual interaction of spins in these multi-atomic systems.

Besides magnetism arising from local magnetic moments, the holistic distribution of the plural elements will also induce the HEX family to present emergent transport behavior. As is known, Anderson localization describes a kind of metal-insulator transition (MIT) where an ordered lattice with uncorrelated random potentials would eventually trap the itinerant electrons[37,38]. The concept of HEX perfectly matches the above criterion. $MS_2$ ($M$=Ti, V, Nb, Ta) are already known to be semimetals or degenerate semiconductors with metallic behavior[39–41]. By imbuing the $M$ site with high entropy, we could easily achieve MIT in $(Ti,V,Cr,Nb,Ta)S_2$ system (**Figure 3**g). When the mobility edge extends through the Fermi level, all the electrons are localized and migrate in the variable-range hopping manner (VRH, **Figure 3**h). The change from a 2D VRH to a $T^{-1/2}$ dependence in $(Ti,V,Cr,Nb,Ta)S_2$ indicates the opening of a Coulomb gap (known as Efros–Shklovskii gap[42]). We point out the VRH has been reported in several 2D systems where the starting point is already a semiconductor (such as $MoS_2$), or the MIT is not dominated by Anderson mechanism (such as $Ti_{1-x}Pt_xSe_2$). Thus far, researches of disorder-induced itinerant to localized transition are mainly focusing on deposited thin films like $Nb_xSi_{1-x}$[43]. The intrinsic configuration of HEX provides a unique platform with abundant candidates to investigate Mott-Anderson MIT. We further investigate the thickness dependent resistivity in **Figure 3**i. Interestingly, a continuous decrease of the resistivity of $(Ti,V,Cr,Nb,Ta)S_2$ can be observed with thinning down to monolayer, which is distinct from the case in other conventional vdW materials. The reduction of interlayer scattering may be responsible for this counterintuitive behavior and a deeper explanation is not clear at present.

HEA has already proved to boost the corrosion resistance[13,15]. It is interesting to see whether this property could be inherited in 2D systems. We choose acid ($HNO_3$), base (NaOH) and organic solution (butylamine mixed in tetrahydrofuran, charged with $O_2$ and CO) as the corrosion reagents. Detailed reaction procedure can be found in Experimental and SI. To include the influence of morphology of different samples, dissolved contents of Se have been calculated in unit per square meter (surface area and raw data for each compound are summarized in **Table S1**). As shown in **Figure 3**j-l, $HESe_2$ presents superior robustness against all its individual counterparts. The dramatic enhancement of corrosion resistance highlights the merit of high entropy. It also worth mentioning that the choice of organic solution is based on the standard procedure of the synthesis of substituted urea, where homogeneous Se is used as the catalyst[44]. Combining the low solubility and high robustness together with the high specific surface area of its 2D nature, HEX may become a promising direction for the design of high-efficient and durable heterogeneous catalysts. Further investigation of its catalytic properties is underway in our lab.

## Conclusion

In this work, we combined the concept of high entropy alloy and vdW materials into a new material category, i.e., HEX, and showcased the physical properties of thus-synthesized dichalcogenides, halides or phosphorus trisulfides. The introducing of multicomponent facilitates the appearance of superconductivity, enriches its magnetic behavior and realizes the Anderson-type MIT. This system also exhibits noticeable enhancement in corrosion resistance which may be a promising route for the design of heterogeneous catalysts.

Needless to say, the marriage of HEA and vdW materials brings us more degrees of freedom and expands the territory of both. Some characteristics of the new HEX materials have been delineated in this work. Considering its infinite possibility (multivariate combination and various properties), we think the intricate coupling of their individual properties also deserve the focus of research community, which is beyond the scope of this work but not the promises of this material. The concept of high entropy could also be introduced to other low dimensional material systems such as (quasi-) 1D materials and MXene phase. Here, we offer an extra rough estimate of the influence of dimension reduction to the entropy of the HEX system: Suppose the number of microstates Ω in each spatial dimension is identical, then $S_{1D}=k_B\ln\Omega$. $S_{2D}=k_B\ln(\Omega*\Omega)=2k_B\ln\Omega$ and $S_{3D}=k_B\ln(\Omega*\Omega*\Omega)=3k_B\ln\Omega$ for 1D, 2D and 3D systems, respectively. Despite the 33% loss of entropy caused by dimension reduction from 3D to 2D, the loosening limitation of anion choice (for example HESSe in **Table 1**), random distribution of the intercalated ionic/molecular species, and the intrinsic fluctuation of the two-dimensional layered structure will somewhat compensate this entropy deficit and bring it with fantastic and unexpected properties. Indeed, the discovery of HEX materials will extend the investigation of 2D materials from limited combinations to unbounded possibilities.

**Table 1 | The synthesized HEX materials and their physical properties**

| Composition (nominal/actual) | Space group Lattice parameters (Å) | Crystal size (mm) | Growth condition | Physical properties |
|---|---|---|---|---|
| (Ti,V,Cr,Nb,Ta)$S_2$<br>$Ti_{0.19}V_{0.2}Cr_{0.14}Nb_{0.24}Ta_{0.26}S_2$ | P-3m1 (164)<br>$a$=$b$=3.3704(2); $c$=5.7676(9) | 5*6*0.2 | (p) 1050 °C, 24 h, quench<br>(s) iodine, 1050 °C, 30 d, quench | Paramagnetic<br>Anderson insulator |
| (Ti,V,Cr,Nb,Ta)$Se_2$<br>$Ti_{0.2}V_{0.21}Cr_{0.21}Nb_{0.19}Ta_{0.2}Se_2$ | P-3m1<br>$a$=$b$=3.5037(5); $c$=6.0620(6) | 2*2.5*0.2 | (p) 1000 °C, 24 h, quench<br>(s) iodine/Se, 1000 °C, 14 d, quench | Paramagnetic<br>Anderson insulator |
| (Ti,Zr,Hf,Nb)$Se_2$<br>$Ti_{0.29}Zr_{0.21}Hf_{0.22}Nb_{0.24}Se_2$ | P-3m1<br>$a$=$b$=3.6311(2); $c$=6.1280(3) | 2*1.5*0.1 | (p) 1200 °C, 24 h, quench<br>(s) iodine, 1000 °C, 14 d, quench | Paramagnetic |
| $(Ti,V,Cr,Nb)_{0.8}(Fe,Mn)_{0.2}Se_2$<br>$Ti_{0.21}V_{0.22}Cr_{0.23}Nb_{0.22}Fe_{0.08}Mn_{0.09}Se_2$ | P-3m1<br>$a$=$b$=3.5133(7); $c$=6.0920(2) | Polycrystalline | 1000 °C, 48 h, quench | Spin glass<br>$T_g$=16 K |
| (Ti,V,Cr,Nb,Ta)SSe<br>$Ti_{0.16}V_{0.15}Cr_{0.15}Nb_{0.24}Ta_{0.17}S_{1.02}Se_{0.98}$ | P-3m1<br>$a$=$b$=3.435(1); $c$=5.945(4) | Polycrystalline | 1000 ℃, 24 h, quench | Spin glass<br>$T_g$=3.5 K |
| (Ti,V,Zr,Nb,Hf)$Te_2$<br>$Ti_{0.15}V_{0.22}Zr_{0.2}Nb_{0.18}Hf_{0.17}Te_2$ | P-3m1<br>$a$=$b$=3.8006(3); $c$=6.6014(2) | 1*1*0.2 | (p) 1000 °C, 24 h, quench<br>(s) iodine, 1000 °C, 14 d, quench | Paramagnetic<br>Metallic |
| $(Co,Au)_{0.2}(Rh,Ir,Pd,Pt)_{0.8}Te_2$<br>$Co_{0.03}Au_{0.06}Rh_{0.23}Ir_{0.24}Pd_{0.16}Pt_{0.28}Te_2$ | P-3m1<br>$a$=$b$=3.9827(1); $c$=5.2601(2) | Polycrystalline | 750 °C, 30 h, quench | Superconductor<br>$T_c$=3.6 K($T_c^{onset}$=4.5K) |
| $Co_{0.1}(Rh,Ir,Pd,Pt)_{0.9}Te_2$<br>$Co_{0.03}Rh_{0.28}Ir_{0.25}Pd_{0.19}Pt_{0.28}Te_2$ | P-3m1<br>$a$=$b$=3.9796(1); $c$=5.2933(1) | Polycrystalline | 750 °C, 24 h, quench | Superconductor<br>$T_c$ = 2.5K |
| (Co,Rh,Ir,Pd,Pt)$Te_2$<br>$Co_{0.19}Rh_{0.23}Ir_{0.2}Pd_{0.15}Pt_{0.18}Te_2$ | P-3m1<br>$a$=$b$=3.9413(1); $c$=5.2967(1) | Polycrystalline | 750 °C, 24 h, quench | Metallic |
| (Mn,Fe,Co,Ni)$PS_3$<br>$Mn_{0.25}Fe_{0.25}Co_{0.25}Ni_{0.27}P_{1.04}S_3$ | C2/m (12)<br>$a$=5.9321(3); $b$=10.2737(5); $c$=6.7061(3) | 3*7*3 | (p) 650 °C, 48 h, quench<br>(s) iodine, 650 °C, 18 d, quench | Antiferromagnetic & Spin glass state, semiconductor<br>$T_N$=70K, $T_{g1}$=35K, $T_{g2}$=56K |
| (Zn,Mn,Fe,Co,Ni)$PS_3$<br>$Zn_{0.29}Mn_{0.14}Fe_{0.17}Co_{0.18}Ni_{0.24}PS_{2.61}$ | C2/m<br>$a$=5.936(1); $b$=10.297(2); $c$=6.713(1) | 1*1*0.1 | (p) 610 °C, 72 h, furnace cooling<br>(s) iodine, 610 °C, 2 d, furnace cooling | Spin glass, semiconductor<br>$T_g$= 30 K, $T_{kink}$= 120 K |
| (Mg,Mn,Fe,Co,Ni)$PS_3$<br>$Mg_{0.19}Mn_{0.18}Fe_{0.19}Co_{0.26}Ni_{0.28}P_{1.08}S_3$ | C2/m<br>$a$=5.953(1); $b$=10.372(3); $c$=6.7394(6) | 1*1*0.1 | (p) 610 °C, 72 h, furnace cooling<br>(s) iodine, 610 °C, 2 d, furnace cooling | Multikinks in MT at 8 K, 42 K, 60 K and 120 K.<br>semiconductor |
| (V,Mn,Fe,Co,Ni)$PS_3$<br>$V_{0.16}Mn_{0.18}Fe_{0.21}Co_{0.23}Ni_{0.24}PS_{2.62}$ | C2/m<br>$a$=5.9431(8); $b$=10.246(2); $c$=6.7080(7) | 1*1*0.1 | (p) 610 °C, 72 h, furnace cooling<br>(s) iodine, 610 °C, 2 d, furnace cooling | Spin glass, semiconductor<br>$T_g$= 37 K, $T_{kink}$ = 150 K |
| (V,Mn,Fe,Co,Ni)$Cl_2$<br>$V_{0.18}Mn_{0.19}Fe_{0.21}Co_{0.2}Ni_{0.18}Cl_2$ | R-3m (166)<br>$a$=$b$=3.5839(2); $c$=17.475(2) | Polycrystalline | 650 °C, 24 h, quench | Antiferromagnetic & Spin glass state, insulator<br>$T_N$= 14.5 K, $T_{g1}$= 9.5 K |
| (Mn,Fe,Co,Ni)$Cl_2$<br>$Mn_{0.26}Fe_{0.25}Co_{0.24}Ni_{0.24}Cl_2$ | R-3m<br>$a$=$b$=3.5782(5); $c$=17.467(6) | Polycrystalline | 650 °C, 24 h, quench | Antiferromagnetic, insulator<br>$T_N$= 15 K |
| (V,Mn,Fe,Co)$Cl_2$<br>$V_{0.19}Mn_{0.27}Fe_{0.27}Co_{0.24}Cl_2$ | R-3m<br>$a$=$b$=3.6177(2); $c$=17.507(2) | Polycrystalline | 650 °C, 24 h, quench | Spin glass, insulator<br>$T_{g1}$ = 7 K |
| (V,Cr,Mn,Fe,Co)$I_2$ | P-3m1 | 10*5*0.2 | (p) 650 °C, 24 h, quench | Spin glass, insulator |

| | | | |
|---|---|---|---|
| $V_{0.23}Cr_{0.18}Mn_{0.19}Fe_{0.19}Co_{0.17}I_2$ (p) | $a=b=4.029(5)$; $c=6.734(1)$ | (s) iodine, 750 °C, 14 d, quench | $T_{g1} = 7$ K |
| $V_{0.15}Cr_{0.12}Mn_{0.24}Fe_{0.26}Co_{0.22}I_2$ (s) | | | |

Without explicit specification of the ratio of individual elements, the total stoichiometric proportion of metallic parts compared with anions are indicated by the bracket. The nominal ratio is equally partitioned for each of the elements, which is consistent with the initial synthesis proportion. (p) and (s) are short for polycrystal and single crystal, respectively. The Rietveld refinements, EPMA/EDX mapping, and basic physical characterizations are included in Supplementary Information.

**ACKNOWLEDGMENTS:**

We thank Yuji Kondo for the help of EPMA measurements, Prof. Shiyan Li and Boqin Song from Fudan University for measuring the thickness dependent resistivity, Prof. Guijun Ma and Prof. Tao Li for fruitful discussions. This project was supported by the MEXT Element Strategy Initiative to form Core Research Center and partially supported by CℏEM (02161943) and Analytical Instrumentation Center (SPST-AIC10112914), SPST, ShanghaiTech University. Y.Q. acknowledges the support by the National Key R&D Program of China (Grant No. 2018YFA0704300), the National Natural Science Foundation of China (Grant No. U1932217 and 11974246), the Natural Science Foundation of Shanghai (Grant No. 19ZR1477300) and the Science and Technology Commission of Shanghai Municipality (19JC1413900).

**SUPPORTING INFORMATION**

The Supporting Information is available free of charge on the ACS Publications website at DOI:XXX

X-ray diffraction, EDX mapping, and basic transport and magnetization measurements of the 17 HEXs listed in **Table 1** (Figure S1-Figure S44); optical image of monolayer and few layers $HES_2$ and the fabricated device (Figure S45-46); XPS measurements of raw, monovalent (K) and divalent (Ba) doped $(Ti,V,Cr,Nb,Ta)S_2$ (Figure S47); corrosion resistance and surface area (Table S1).

**AUTHOR INFORMATION:**

*t-ying@mces.titech.ac.jp

*qiyp@shanghaitech.edu.cn

*hosono@mces.titech.ac.jp

## Table of Contents

The multi-degrees of freedom inherited from high entropy alloy and van der Waals materials are intermingled together into a brand-new material category, namely HEX materials. Single crystals with homogeneous element distribution can be easily obtained, exfoliated and intercalated. With infinite combinations, HEX materials show enhanced corrosion resistance and versatile physical properties, covering superconductivity, metallic, semiconducting and insulating states.

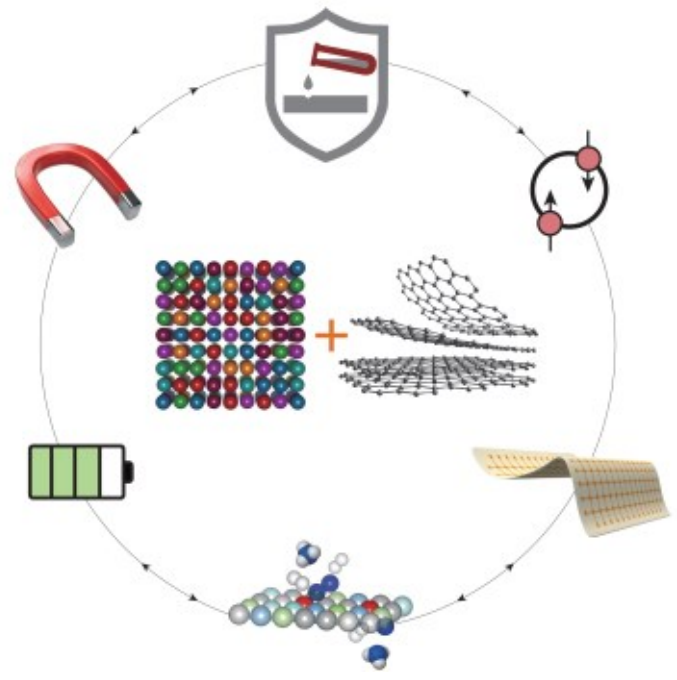